\documentclass{ws-procs9x6}

\begin{document}

\title{Energy Dependence of $pp$ and $p$-carbon\\ CNI Analyzing Power\footnote{\uppercase{T}his work is authored under contract number \uppercase{DE-AC02-98CH10886} with the \uppercase{U.S. D}epartment of \uppercase{E}nergy.}}

\author{T.~L. Trueman
\address{Physics Department, \\
Brookhaven National Laboratory,\\
Upton, NY 11973, USA\\
E-mail:trueman@bnl.gov}}
\maketitle

\abstracts{
The method described in my RHIC Spin Note (hep-ph/305085) is applied to recently reported data from RHIC experiments in order to obtain values for the spin-flip Regge couplings. The data comes from both 100 GeV/c proton elastic scattering on a carbon target and on the recently commissioned polarized hydrogen gas jet target. These couplings are used to predict the analyzing power for proton-carbon scattering at the top RHIC fixed target energy of 250 GeV/c and for $pp$ scattering at RHIC collider energy. }

\section{Background}
The Coulomb-nuclear interference induced asymmetry  in proton elastic scattering has been seen to be a very sensitive and practical polarimeter for high energy proton beams\cite{Jinnouchi,Okada}. The main uncertainty in calculating this analyzing power is the unknown hadronic-spin flip amplitude, which we characterize by the quantity $\tau$, the ratio of the single-flip to the non-flip proton scattering amplitude. $\tau$ cannot be calculated with the requisite precision and so it is necessary to calibrate the polarimeter. For a carbon target this is possible at AGS energy, about 24 GeV/c, where other means of determining the beam polarization are available, but it is not possible at higher energy. The goal of my RHIC Note\cite{note} is to calculate the analyzing power at higher RHIC energy when the polarization is known at only the injection energy.  This work relies heavily  on earlier work\cite{KT} which shows, among other things, that it is  quite reasonable to assume that $\tau$ for $pC$ is equal to the $I=0$ part of $\tau$ for $pp$ and that its variation with $t$ over the CNI region can be neglected.

Its variation with $s$ needs to be determined.\ (We will often use the beam momentum $p$ and the proton-proton c.m.\ energy squared $s= 2 m^2 + 2 m \sqrt{p^2 + m^2} $ interchangeably.) To address this, I have taken a simple model that has been successfully used for unpolarized scattering over a wide energy range, namely the Regge pole model. For this work we chose the simple Regge pole model whose parameters are determined by \cite{Cudell}: three poles are found to contribute significantly: the pomeron, a $C=-1$ Regge pole, mainly the $\omega$, and a $C=+1$ Regge pole, mainly the $f$. We extend the model to polarized $pC$ scattering by neglecting the small $I=1$ contributions to the $pp$ unpolarized scattering data used in the fit and we introduce the spin-dependence through three real parameters $\tau_P, \tau_f$ and $\tau_\omega $ that give the ratio of the spin-flip to non-flip Regge couplings for the pomeron, the $f$ and the $\omega$. They do not depend on the momentum of the beam. Until recently the only determination of the $pC$ analyzing power at high energy with a beam of know polarization was E950 at the AGS with beam momentum of 21.7 GeV/c.  $\tau(21.7)$ determined by this experiment is found to be non-zero and to have a significant imaginary part. The real and imaginary parts of $\tau(21.7)$ can be used with our model to determine two relations between the three real Regge spin-flip couplings in our model. With one more relation, the model would be completely determined. 

A good measurement of the shape of the curve at a different energy with a polarized beam of unknown polarization can provide such a measurement because shape of $A_N$, defined by,  
\begin{equation} 
S(s)= \frac{Im(\tau(s))}{\kappa/2 - Re(\tau(s))}
\end{equation}
can be determined independently of $P$.  We fit
\begin{equation}
\frac{A_N(s,t,\tau(s))}{A_N(s,t,0)} =\frac{2}{\kappa}\{\kappa/2 - Re(\tau(s)) + f(s,t) Im(\tau(s))\}
\end{equation}
where $A_N(s,t,0)$ is the "pure" CNI" analyzing power in the absence of any hadronic spin flip, $\kappa = 1.79$ and $f(s,t)$ is a function that can be calculated using Glauber techniques without any spin-dependent information \cite{KT}. This is all valid to terms linear in $\tau(p)$. At the Spin2002 preliminary values for $A_N$ at 100 GeV/c were reported \cite{Kurita} and in \cite{note} we used them to determine S(100~GeV/c). This gives us a third relation among the Regge couplings and so we can calculate the analyzing power at 100 GeV/c without measuring the polarization at that energy. The prediction is $\tau(100) = -0.130 -0.053 i$ with an error of about $\pm 0.22$ on the real part but only $\pm .012$ on the imaginary part.
\section{New pC results}
At this meeting new results with much smaller errors and known beam polarization (as measured by the gas jet) at 100 GeV were reported\cite{Jinnouchi}. By fitting this data one finds $\tau(100)= -0.017 -0.049 i$  within errors of the above prediction. The analyzing power is predicted by this model  to be uniformly about 
10 \% above the data (Note that for  the data used  only statistical errors are now available. Systematic errors will be available soon.)

Since the errors on the new data are much smaller, lets  turn the process around and use the $\tau$-value determined by the new 100 GeV/c data and use just the shape from E950. This leads to a determination of the Regge couplings with much smaller errors than before \cite{note}. See Fig.1.
\begin{figure}[ht]
\centerline{\epsfxsize=4.1in\epsfbox{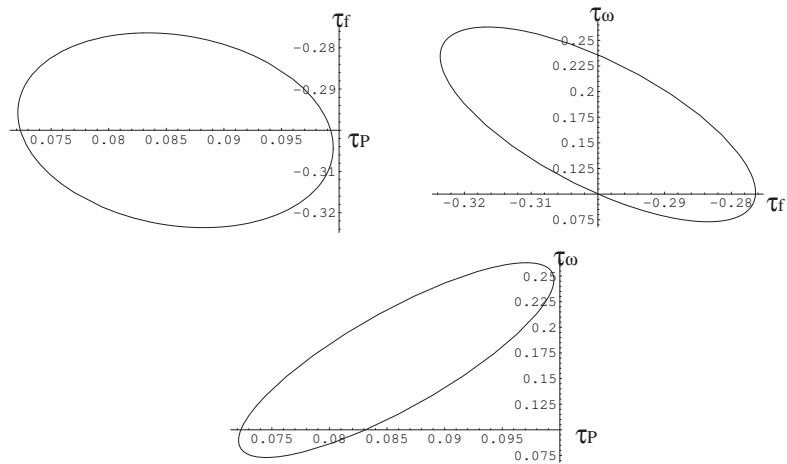} }  
\caption{1-sigma error ellipses for the three Regge spin-flip couplings}
\end{figure}

These can be used to predict the $pC$ analyzing power at 250 GeV/c shown in Fig.2. The thickness of the curve represents the statistical errors of this determination.
\begin{figure}[ht]
\epsfxsize=10cm   \centerline{\epsfxsize=4.1in\epsfbox{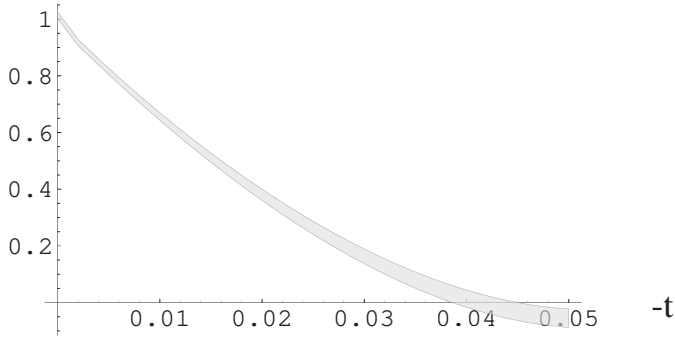}}   
\caption{Predicted $A_N(s,t,\tau(s))/A_N(s,t,0)$ at 250 GeV/c}
\end{figure}

\section{$pp$ CNI}

In order to extend the model and make predictions for high energy $pp$ scattering asymmetries we need information about the $I=1$ Regge poles.  The pomeron coupling is the same as already determined. We need in addition the $\rho$ and the $a_2$.  If we assume that they are, respectively, degenerate with the $\omega$ and $f$, the pairs will enter the fit as only two Regge poles, one with  $C=-1$ and one with $C=+1$.
We are now blessed with the $p$-jet data at 100 GeV/c and so have $A_N(100)$ \cite{Okada}. So the real and imaginary parts of the corresponding spin-flip factor can be used to determine the two new constants, $\tau_-$ and $\tau_+$, and the model is completely determined. The values for the coupling are
\begin{eqnarray}
\tau_P=0.09 \nonumber \\
\tau_+= -0.32 \nonumber \\
\tau_-=1.06 \nonumber
\end{eqnarray}
We see from this that the $a_2$ coupling is rather small but the $\rho$ coupling is enormous. This was to be expected \cite{berger}.

These couplings can be used to predict the analyzing power in colliding beams of protons at high energy, such as in the recent $pp2pp$ experiment at RHIC. The prediction is  $\tau(s=200^2)= 0.08 -0.007 i$; the significant non-zero part has an error of about 2\% due to error in the pomeron coupling. 
The preliminary data reported to this conference \cite{alexeev} lies significantly above this prediction;  this discrepancy will have to be addressed.

\end{document}